# Magnetic field induced dielectric relaxation in the strain glass state of $Pr_{0.6}Ca_{0.4}MnO_3$

K Devi Chandrasekhar[1], A K Das[1] and A Venimadhav[2,a)]

[1]Department of Physics & Meteorology, Indian Institute of Technology, Kharagpur, West Bengal 721302, India

[2] Cryogenic Engineering Centre, Indian Institute of Technology, Kharagpur, West Bengal 721302, India

We present the dielectric and magnetodielectric properties of $Pr_{0.6}Ca_{0.4}MnO_3$ polycrystalline sample. Dielectric permittivity ($\varepsilon'$) (and $d\varepsilon'/dT$) portrays the charge order and other magnetic transitions observed in the magnetization measurement. Dielectric study has revealed a relaxation corresponding to ordering of polarons ~ 60 K that follows Arrhenius behaviour both in the presence and absence of magnetic field and another relaxation was noticed ~ 30 K only under a critical magnetic field (3.2 T) that shows critical slow down of electronic charges obeying power law. Further, the magnetic field induced relaxation shifts to low temperatures with the increase of magnetic field. The observed field induced dielectric relaxation below the reentered charge ordered state is linked with the rapid motions of boundaries of the coexisting phases towards the martensite phase transformation.

**Introduction**

A close competition among various magnetic and electronic ground states in cuprates, cobaltates and manganites results magnetoelectronic inhomogeneity [1-4]. In particular, spatial coexistence of various phases without chemical inhomogeneity has been experimentally observed in both electron and hole doped manganites [2, 5, 6]. The scenario of electron conduction with core spins ordered ferromagnetically, giving rise to double-exchange ferromagnetism is opposed by strong antiferromagnetic Heisenberg superexchange between the core spins established by orbital-related mechanisms. The relative strength of the competing interactions is governed by chemical pressure and doping [5]. The phase separation (PS) resolves the competition by dividing the system into regions of different competing interactions; a combination of martensitic strain due to coexisting phases and the effects of strong electron correlations appear to manifest the PS [7-9]. The existence of PS has been verified experimentally by several techniques such as electron diffraction, neutron diffraction measurements, Lorentz electron microscopy, high resolution electron microscopy and optical techniques [10-14]. The PS mechanism is sensitive to external stimuli such as electromagnetic radiation, magnetic and electric fields; percolation takes place with conversion of antiferromagnetic (and ferromagnetic) insulating state to ferromagnetic metallic (FMM) state [15-18]. A similarity between phase separated manganites and magnetic glassy systems has been suggested and such a scenario is evident due to frustration of the FMM as well as the charge ordered (CO) insulating states at the phase boundary [9]. A number of studies pertaining to the electronic glassy dynamics of the phase separated region were demonstrated through large 1/f noice and resistive memory experiments [3, 19-25].

More studies were done on $(La_{1-x}Pr_x)_{1-y}Ca_yMnO_3$ and $Pr_{1-x}Ca_xMnO_3$ systems due to their large CMR effect at moderate magnetic fields [16, 23]. At low temperatures, the strongly interacting phase separated regions undergoes a cooperative random freezing from dynamic to static phase separated state, indexed by the re-entrant CO transition and this has been termed as a strain liquid to strain glass transition as evidenced from the thermal measurements [7]. In $Pr_{1-x}Ca_xMnO_3$ series, for smaller x, ferromagnetic insulating state appears at low temperature, but for x ~ 0.3-0.5 the canted antiferromagnetic insulating (CAFMI) state is more prominent [16]. Recently, the dynamic characteristics of the charge carriers in thin films of phase separated manganites have been probed using dielectric spectroscopy [26]. The anomaly at CO transition and melting of ordered polarons in $Pr_{1-x}Ca_xMnO_3$ (x = 0.3, 0.33) at low temperatures (~ 60 K) were substantiated from dielectric measurements [27, 28]. In the present work, we report the observation of magnetic field induced glassy dielectric behaviour in $Pr_{0.6}Ca_{0.4}MnO_3$ (PCMO) sample at the metal-insulator phase boundary below the reenter CO transition. Rapid motion of boundaries of the coexisting phases towards the martensite phase transformation seems to induce the dielectric relaxation.

**Experiment**

Polycrystalline $Pr_{0.6}Ca_{0.4}MnO_3$ sample are prepared by sol-gel method and the final sintering was done at 1300 °C. The High resolution X-ray diffraction (HRXRD) from Panalytical with Cu Kα radiation, was used to determine the structure and Rietveld refinement was done using FullProf suite software. The crystal structure was refined using orthorhombic Pbnm space group with lattice parameter of a = 5.415 Å, b = 5.438 Å and c = 7.664 Å [29]. For electrical resistivity a standard four probe technique has been employed from room temperature to 10 K. To avoid the Joule heating effects measurements are performed with very low current of magnitude 1μA (using keithely-6221 current source and keithely-2002 nanovoltmeter). The static magnetization measurements were performed using Quantum Design VSM-SQUID magnetometer. Dielectric measurements were performed in the temperature range 5 K-300 K using the Agilent HP4192A impedance analyzer from 40Hz-1MHz. High purity silver paste contacts were applied on both faces of the pellet to make parallel capacitance geometry. The low ac excitation voltage (20 mV) has been employed to maintain the linear capacitance voltage characteristics. The temperature and magnetic field dependent transport and dielectric measurements were performed in the JANIS cryogenfree superconducting system with closed cycle helium refrigeration.

**Results and Discussion**

The temperature dependent magnetic measurements (M-T) has been performed in the zero field cool (ZFC) and field cool (FC) protocol with an applied magnetic field of 100 Oe as shown in the inset of fig. 1(a). The magnetization in M-T curve exhibits a peak near 237 K corresponding to charge ordering and successive magnetic transitions at 170, 120 and 42 K corresponding to antiferromagnetic (AFM), ferromagnetic (FM) and canted ferromagnetic transition (CAFM) is observed [16, 30]. The presences of multiple magnetic transitions indicate the inhomogeneous magnetic nature due to coexisting phases of FMM and COAFMI phases. In order to reveal the PS nature, we have performed the ZFC measurements for different applied fields as shown in fig. 1(a). For magnetic field above 3.2 T the ZFC, curve exhibits the step like increase and decrease of magnetization between 28 K - 80 K and this magnetic field induced PS behaviour matches with that of $Pr_{0.6}Ca_{0.4}MnO_3$ single crystals [31]. For higher magnetic fields, the AFM transition at 170 K shifts to the low temperatures and disappears above 6 T, where as the CO transition exists even up to 7 T indicating the robust nature of the CO state. The low temperature CAF transition disappears below 3 T magnetic field. The nature of the M-T curves under high fields (> 3.5 T) resembles to that of $La_{0.215}Pr_{0.41}Ca_{3/8}MnO_3$ (LPCMO) systems, where M-T curves were interpreted as a consequence of different kinds of PS zones. The crossover point in ZFC and FC curves at low temperature can be assigned to the transformation from the static phase separation (SPS) to dynamical phase separation (DPS) region [7,32]. In analogy with La$_{0.215}$Pr$_{0.41}$Ca$_{3/8}$MnO$_3$ system we have distinguish the DPS and SPS region from the ZFC and FC curves (H = 4 T) and are shown in inset of fig. 1(b); the arrows indicate the heating and cooling cycles respectively.

Fig. 1(b) shows the temperature dependent resistivity measurements for different applied magnetic fields under ZFC mode. The zero field resistivity shows the semiconducting behaviour with slope changing around T$_{CO}$; with the increase of magnetic field the insulating state at low temperatures persist for magnetic fields up to 2.2T. At 3.2 T, the resistivity exhibits an insulator to metal behaviour due to melting of COI state below the FM transition and this peak shifts to higher temperature with the increase of magnetic field. A sharp rise in resistivity with magnetic field can be observed near strain liquid to strain glass (30 K) transition due to reentered CO state (inset of fig. 1(b)). This is in contrast to x=0.3 phase where the field induced FMM state is more persistent [16].

The variation of magnetization, magnetoresistance and magnetodielectric effect at 30 K (below the CAFMI transition) in the static PS region and at 50 K in the dynamic PS state are shown in the fig. 2 (a) and (b). At 30 K, magnetization increases slowly with the increase of H and shows an abrupt jump at ~ 3.2 T and it saturates above 6.5 T implying the change of initial AFM state to FM (H$_{AFM\rightarrow FM}$). But reversing the magnetic field, shows a large irreversibility with a hysteresis governed by several metastable states within coexisting magnetic phases [2, 5]. Once after the complete field sweep (7 T→-7 T→7 T) the field induced FM moment partially remains due to kinetic arrest of FM phase within the AFM matrix and this is evident from the little rise of magnetization at very low magnetic fields as shown in the inset of upper panel of fig. 2(a) [33]. At 50 K, above the CAFI transition, the induced FM moment is much small as shown in the upper panel of the fig. 2(b). This kinetic arrest of FM phase in the AFM background resembles in both resistivity and permittivity, where (the lower pane of fig. 2 (a)), the initial state was not reproduced on returning to zero field. The lower panel of fig. 2 (a) shows a decrease of resistivity by five orders of magnitude at 3.2 T and an increase of dielectric permittivity by three orders at the same magnetic field. The colossal dielectric response observed in $Pr_{1-x}Ca_xMnO_3$ (x = 0.3, 0.37) single crystals has been attributed to the dielectric catastrophe at the insulator to metal transition and it is an indication of perolative transport [34].

The temperature dependent dielectric permittivity and the loss measured without magnetic field for different frequencies are shown in the fig. 3(a) and (b) respectively. The dielectric permittivity portrays distinct transitions observed in the magnetic data (a) a small discontinuity in dielectric and corresponding loss tangent at 230 K near CO transition (b) a peak below 200 K that can be related to the onset of AFM transition (c) ~ 60 K the dielectric permittivity decreases very sharply accompanied by a peak in the loss tangent that shifts to low temperatures with the decrease of frequency (fig. 3(b)). The low temperature dielectric relaxation observed ~ 60 K is consistent with other reports on the dielectric behaviour of PCMO where this relaxation has been related to the polaron charge carriers at the localized sites [28, 35]. At low temperatures (below 30 K) the dielectric permittivity shows a saturation value of ~ 30 and this is of the same order of intrinsic dielectric permittivity of perovskite materials [28, 35].

The magnetic field variation of dielectric permittivity for 100 kHz is shown in the fig. 3(c). The dielectric permittivity up to 2.2 T overlaps with the 0 T curve. Here, the polaron relaxation ~ 60K, which we call as relaxation 'A', exhibits a large change in its magnitude with the increase of magnetic field and the relaxation peak shifts to low temperatures (fig. 3(d)) indicating the melting of the localized polarons [28]. With the magnetic field increased to 3.2 T, the dielectric permittivity shows another step like feature below the relaxation 'A' peak. It can be seen more clearly from the inset of fig. 3(c) as peak in dε'/dT with respect to temperature; this peak shifts to low temperatures with the decrease of frequency as shown in the fig. 3(d); this illustrates the relaxation nature of the peak. This we call as relaxation 'B' and it appears at the onset of DPS to SPS (30 K) cross over for a critical magnetic field (H > 3.2T). The dielectric permittivity with the application of 5 T field shows the shifting of the 'B' relaxation to the low temperatures. The appearance magnetic field induced additional relaxation 'B' in PS can be correlated with strain glass to strain liquid cross over, where a sudden growth of FMM clusters within the AFMI background leads to an insulator to metallic transition [7]. At low temperatures due to reentrant CO phase, higher magnetic field is necessary for the percolation to occur; hence the 'B' relaxation shifts to the low temperature for higher field.

The magnetic field induced dynamical behaviour of charge carriers is analyzed by fitting relaxation peak of dielectric loss to thermally induced Arrhenius mechanism. Fig. 4 shows the plot between $\ln\tau$ vs. 1/T for different applied fields for relaxation 'A'. As shown in the figure the relaxation obeys the Arrhenius law given by

$$\tau = \tau_0 \exp\left(\frac{E_a}{k_B T}\right) \qquad (1)$$

where $\tau_0$ is the pre-exponential factor, $k_B$ is the Boltzmann constant and $E_a$ is the activation energy. The linear line indicates fit to the equation (1). The variation of the obtained fitting parameter with respect to applied magnetic field is shown in the inset of fig. 4. As shown in the figure the activation energy remains unchanged up to 2.2 T and it decreases abruptly to 30 meV at 3.2 T, further the relaxation time changes by an order of magnitude. The obtained activation energy and the relaxation time are consistent with the polaron hopping energies [28, 35]. The decrease in activation energy with magnetic field indicates the decrease of polaron hopping barrier [28].

The dynamics of relaxation 'B' appeared for the 3.2 T (in fig. 3 (d)) is examined using Arrhenius mechanism and the relaxation time obtained to be of the order of ~ $10^{-17}$ sec. This is physically unrealistic and the relaxation process seems to follow a complex behaviour. The observed relaxation mechanism has been verified with power law mechanism and is given by

$$\tau = \tau^0 (\frac{T}{T_g} - 1)^{-zv} \qquad (2)$$

where $\tau^0$, zv are the glass temperature and critical exponent respectively. The best fit to power law is shown as solid line in inset of fig. 4 with fitting parameters $\tau^0$ = 1.13x$10^{-8}$sec, $T_g$ = 24.1 K and zv = 6.5. Dielectric data following the power law indicates the glassy critical behaviour of the system. The glassy nature of the 'B' relaxation is apparent as the changeover from strain glass to strain liquid transition is associated with glassy behaviour in magnetic, transport and structural properties. In fact the thermal and magnetic field induced SPS to DPS results different conduction paths like, polaron hopping, metallic, and tunnel conductivity between the FM clusters that may lead to the glassy dynamic behaviour [6, 7, 32].

The large change in dielectric permittivity and resistivity (see fig. 2) at above and below the reentrant CO transition at a critical magnetic field suggests the percolation nature of the transport. However, field induced dielectric relaxation is only observed below reentrant CO transition. A large strain at the interface between FMM and COI phases in the DPS region cooperatively freezes in SPS region leading to strain liquid to strain glass transition [7, 32]. Application of magnetic field in the strain glass state leads to manifestation of a martensitic structural transition with rapid motion of boundaries. Here magnetic phase transformation is coupled to the structural mismatch between the pseudotetragonal CO and pseudocubic FM phases that induces changes in the dynamic properties along with lattice strain [8, 12, 32]. Like in $La_{0.215}Pr_{0.41}Ca_{3/8}MnO_3$, one can expect a strain liquid to glassy behaviour in PCMO below reentrant CO transition. Below this transition, the charge carriers are localized under the AFM background in the form of polaronic solid thus, resulting insulating state. At a critical field (percolation threshold) the competition between the charge ordered AFMI and charge disordered FMM phase becomes more dynamic with fast growing of FM state within the AFMI matrix. The observed relaxation 'B' indicates the martensite like transformation involving large strain due to the relative motion of boundaries between the two phases in the strain glass state, moreover the changeover of strain glass to strain liquid state is associated with the several conduction mechanisms as discussed above and these combined effects manifest the glassy dielectric behaviour. For higher magnetic field the peroration occurs at a lower temperature and hence the relaxation 'B' shifts to low temperature. The critical behaviour of the relaxation in other words indicates the slow dynamics within the PS under the magnetic field.

In summary, dielectric permittivity $\varepsilon'$ (and $d\varepsilon'/dT$) portrays the CO and magnetic transitions in the PCMO system. Dielectric data distinguishes the relaxation due to polarons that follows the usual activation behaviour. A field induced relaxation in the strain glass region has been revealed and it is associated with the rapid motion of boundaries between the competing phases. The manifestation of critical slow down of the carriers is verified by dielectric glassy relaxation behaviour. In a border perspective, dielectric measurements can show the signature of martensite like transformations.

**Acknowledgement:**

We acknowledge M. B. Salamon and S.W. Cheong for helpful discussion and comments on the manuscript. This work was supported DST fast track project the authors also acknowledge the use of SQUID VSM facility set up by IITKGP.

**Figure Captions:**

Fig. 1 (a) Temperature variation ZFC magnetic for different applied magnetic fields, inset shows the temperature variation of ZFC-FC magnetization for 100 Oe. (b) Temperature variation of resistivity for different applied magnetic fields, inset shows the temperature dependent magnetization (ZFC-FC) and resistivity for 4 and 4.3 T respectively.

Fig. 2 (a) Upper panel shows the M-H plot at 30K, inset shows magnified view at low magnetic fields and the lower panel shows the magnetic field variation of resistivity and dielectric permittivity at 30 K. (b) Upper panel shows the M-H plot at 50 K, inset shows magnified view at low magnetic fields and the lower panel shows the magnetic field variation of resistivity and dielectric permittivity at 50 K

Fig. 3(a) Temperature dependent dielectric permittivity ($\varepsilon'$) (b) Tan $\delta$ for different frequencies at H = 0 T. (c) Temperature dependent dielectric permittivity for different applied fields at 100 kHz; inset variation of $d\varepsilon'/dt$ with respect to temperature. (d) Variation of $d\varepsilon'/dt$ with respect to temperature for different frequency at H = 3.2 T.

Fig. 4 Shows the plot between $\ln\tau$ vs. inverse temperature for different magnetic fields and solid lines shows the fit to eq. (1); left inset shows the magnetic field variation of $E_a$ and $\tau_0$, right inset shows the plot between $\tau$ vs. temperature for 3.2 T and solid line indicated the fit to eq. (2).

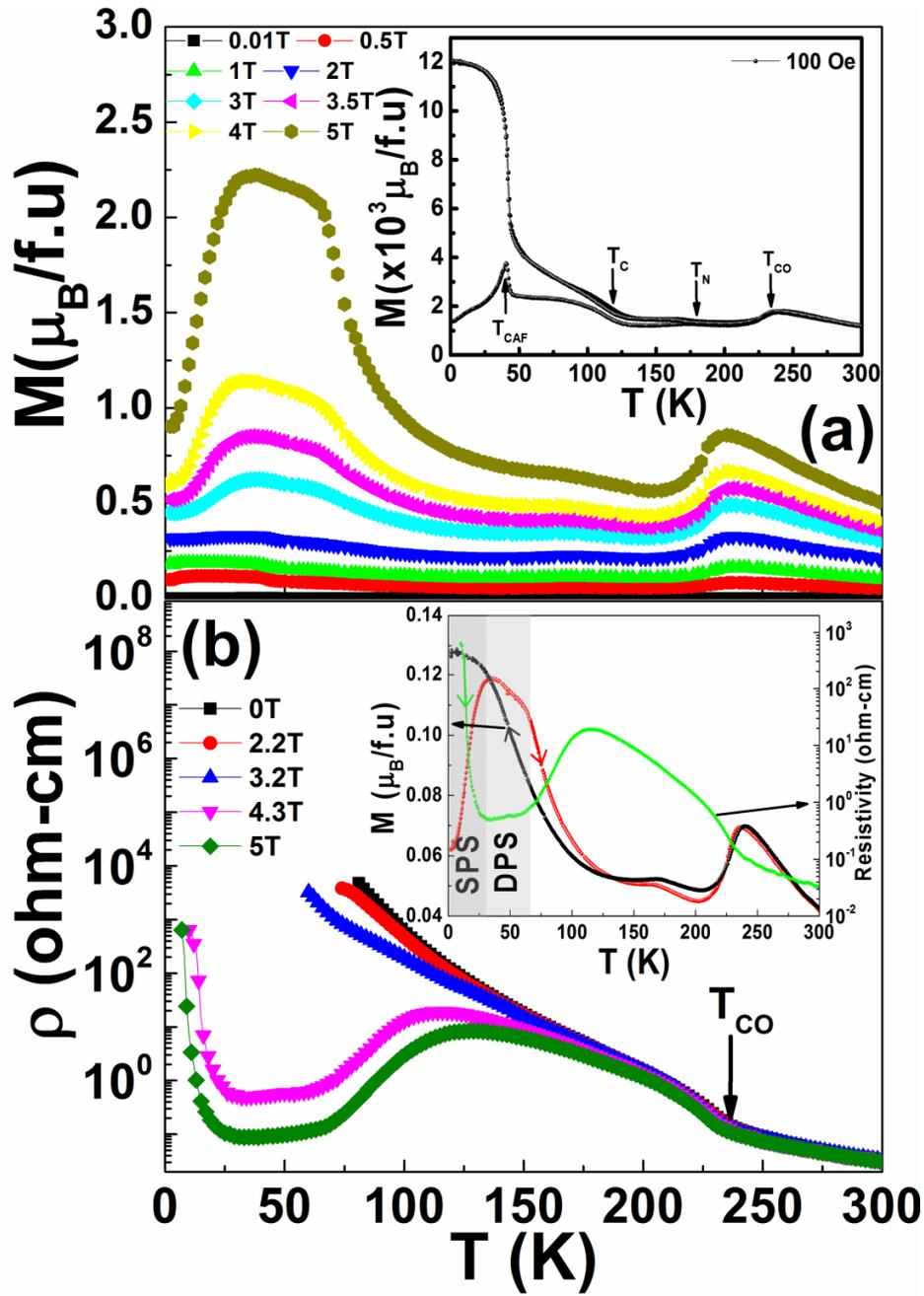

Fig 1

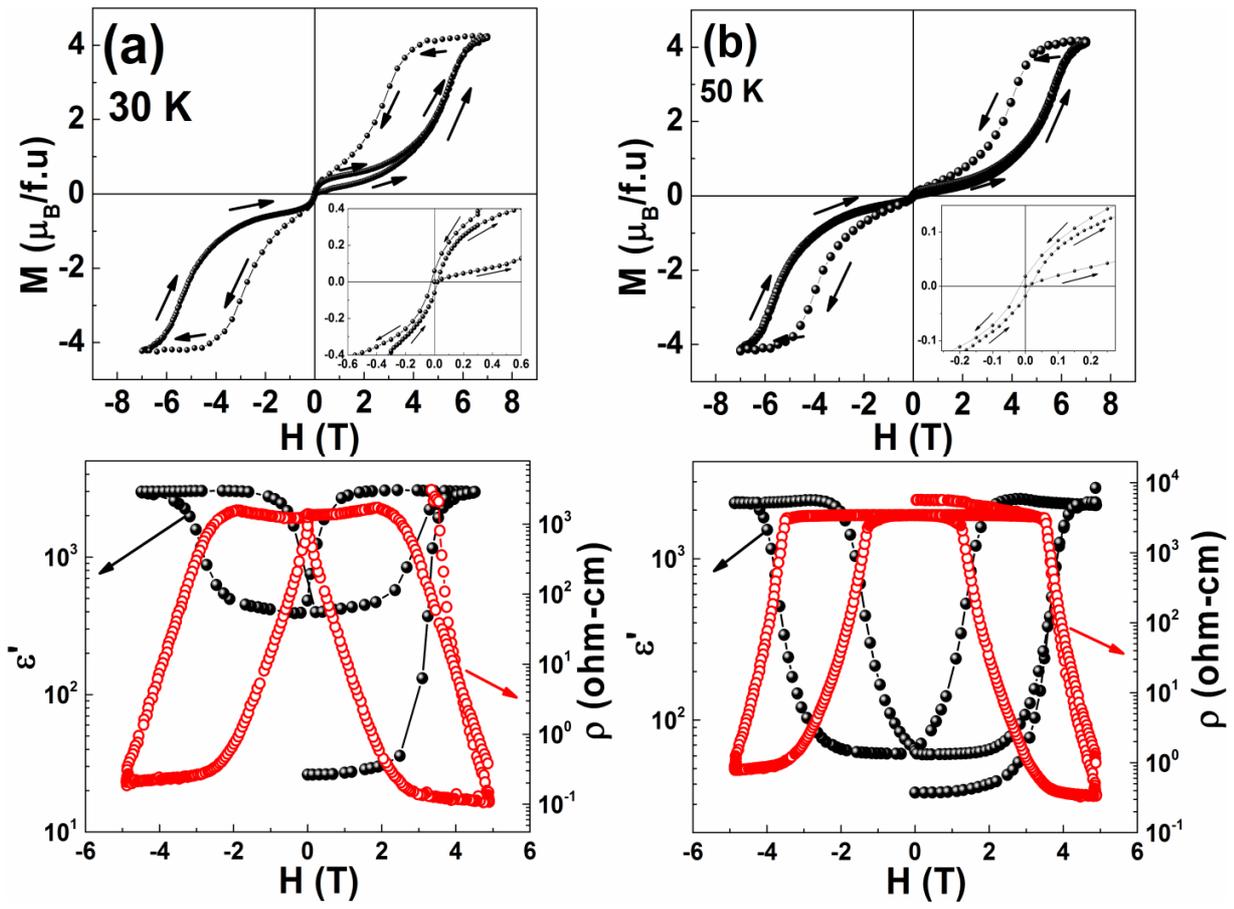

Fig 2

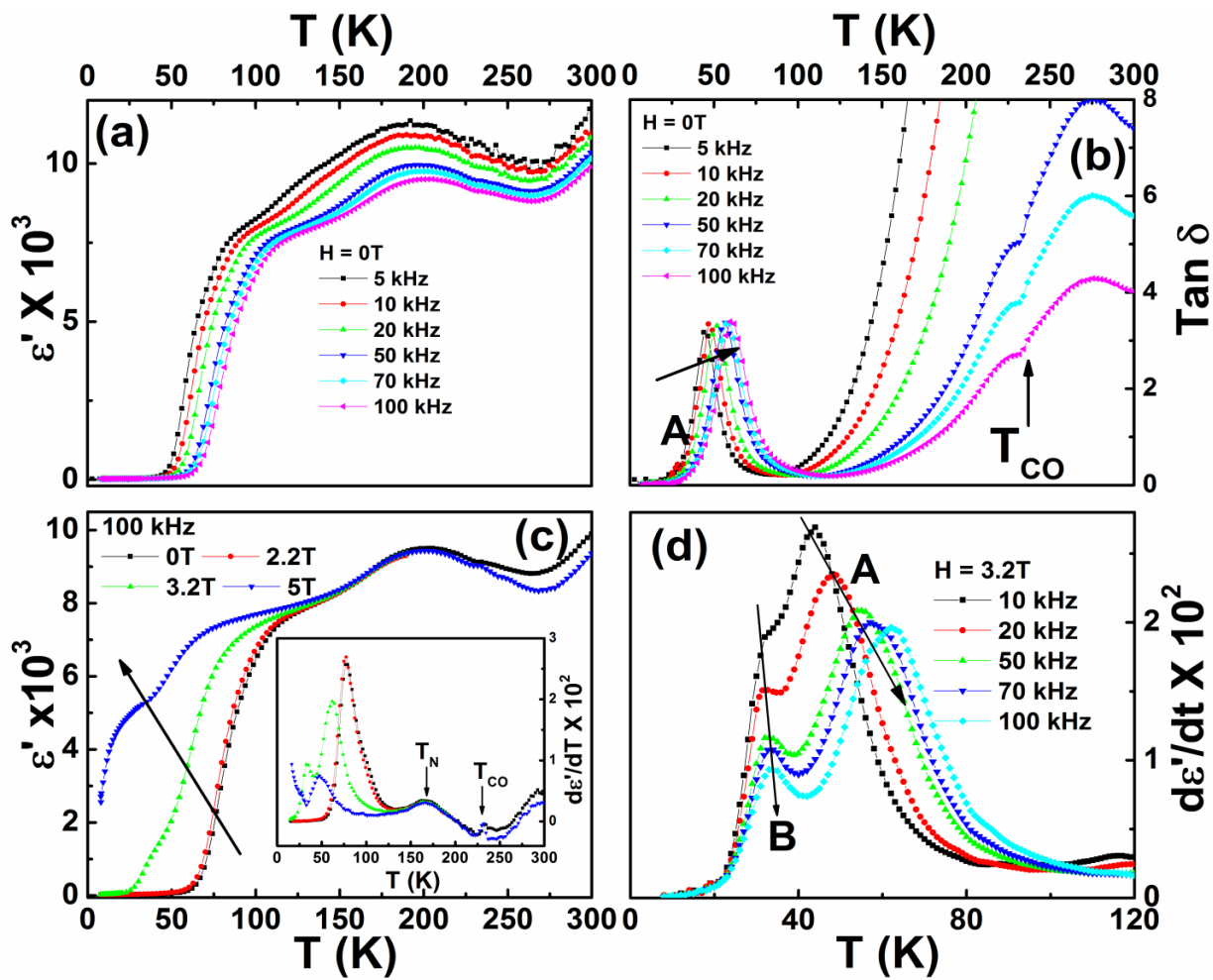

Fig 3

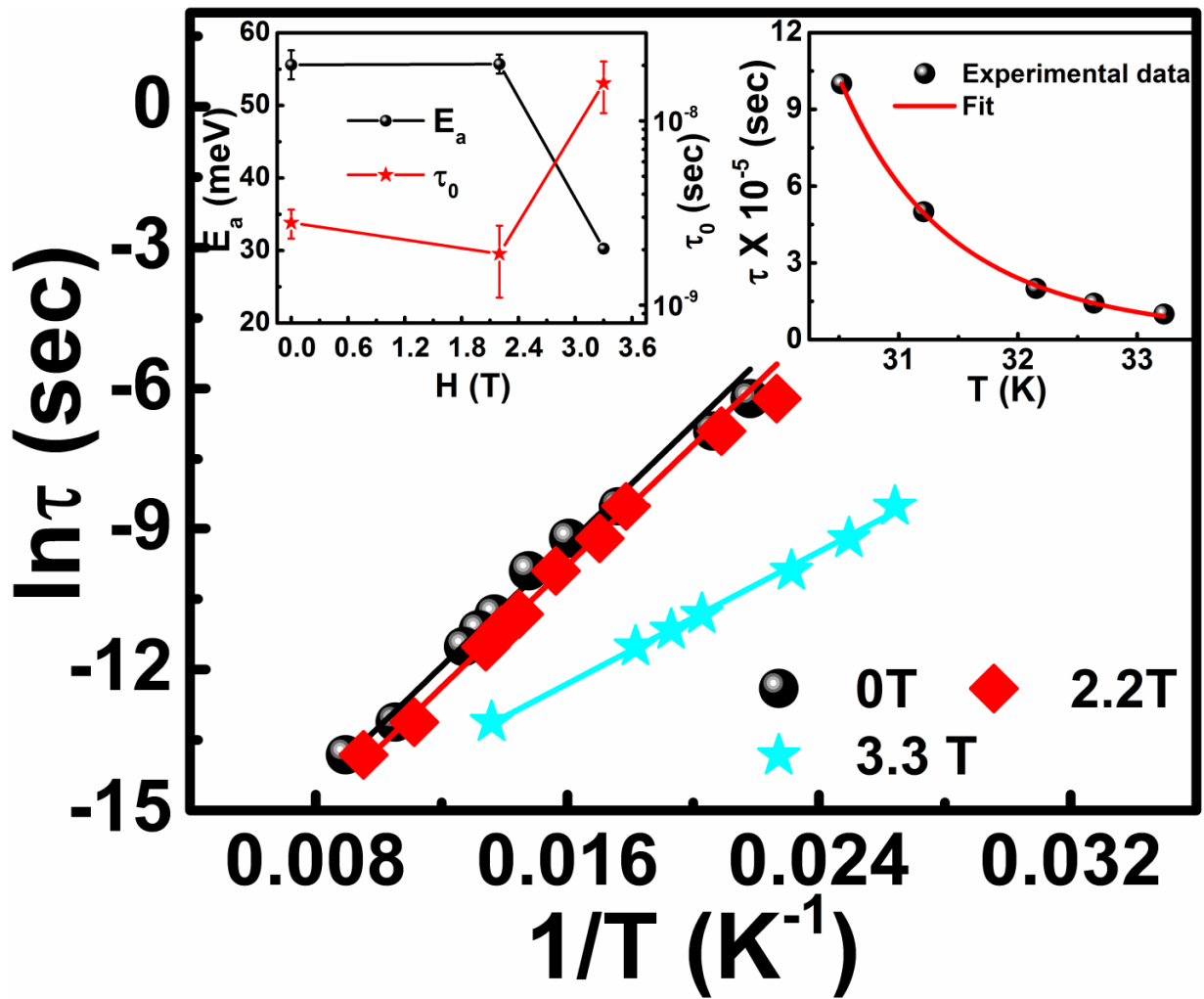

Fig 4